# Influence of the aggregate state on band structure and optical properties of C60 computed with different methods

Amrita Pal[1], Saeid Arabnejad[2], Koichi Yamashita[2], and Sergei Manzhos[1]

[1]*Department of Mechanical Engineering, National University of Singapore, Block EA #07-08, 9 Engineering Drive 1, Singapore 117576, Singapore*

[2]*Department of Chemical System Engineering, School of Engineering, University of Tokyo, 7-3-1, Hongo, Bunkyo-ku, Tokyo 113-8656, Japan*

C60 and C60 based molecules are efficient acceptor and electron transport layers for planar perovskite solar cells. While properties of these *molecules* are well studied by *ab initio* methods, those of *solid* C60, specifically its optical absorption properties, are not. We present a combined Density Functional Theory – Density Functional Tight Binding study of the effect of solid state packing on bandstructure and optical absorption of C60. The valence and conduction band edge energies of solid C60 differ on the order of 0.1 eV from single molecule frontier orbital energies. We show that calculations of optical properties using linear response TD-DFT(B) or the imaginary part of the dielectric constant (dipole approximation) can result in unrealistically large redshift in the presence of intermolecular interactions compared to available experimental data. We show that optical spectra computed from the frequency-dependent real polarizability can better reproduce the effect of C60 aggregation on optical absorption, specifically with a GGA functional, and may be more suited to study effects of molecular aggregation.

# 1 Introduction

C60 based materials are widely used in photoelectrochemical / solar cells such as organic solar cells[1, 2] and planar perovskite solar cells.[3, 4] While in bulk heterojunction cells, addends are used to control solubility and the morphology of the heterojunction (PC60BM is the most widely used example),[5, 6] in planar perovskite cells, bare C60 (and C70) perform well.[7] Planar perovskite cells have now achieved conversion efficiencies about 20%[8] and are most promising given their advantages in fabrication.[9, 10] Therefore, characterization of fullerene layers, both experimental and theoretical / computational, is highly technologically relevant. The main role of these materials is to ensure charge separation at the interface with the donor and electron transport to the electrodes. Their electron affinity / conduction band minimum (CBM) are critical determinants of the efficiency of charge separation. To avoid voltage losses, one strives to achieve a minimal offset from the conduction band minimum (or LUMO, lowest unoccupied molecular orbital, of the corresponding molecules) of the donor at which charge separation is still efficient. Therefore, the precise position of the CBM is important.

The CBM is determined by the energy of the LUMO of the fullerene molecule (conversely, the VBM, valence band maximum, is determined by the energy of the HOMO, highest occupied molecular orbital) and by its interaction with the environment, which is made mostly of other fullerene molecules in the case of solid layers. Single-molecule HOMO/LUMO estimates can be quite accurate both by experiments (e.g. by cyclic voltammetry[11] in solution) and by modeling, as highly accurate *ab initio* methods (using hybrid functionals or even wavefunction-based methods with large basis sets) are feasible for single molecules. In solids, characterization is more difficult. For example, the onset points of the PESA (photoelectron spectroscopy in the air[12]) signal (for HOMO analysis) have rather wide tolerances.[13, 14] Optical bandgaps (often used for LUMO estimates[15, 16] of organic molecules) are also not effective as absorption spectra of fullerenes are not dominated by HOMO→LUMO transitions as they are e.g. in most organic dyes.[17] Highly accurate computational methods are typically also unfeasible in solids. It is therefore important to have computed estimates of the change in LUMO/CBM and HOMO/VBM due to aggregate state to gauge their effect on band alignment in solar cells. While C60 and C70 molecules have been characterized *by ab initio* methods in multiple studies,[17, 18] this has not been the case for solids, specifically as far as optical properties are concerned, with only a handful studies available.[19, 20]

C60 does not possess significant visible absorption but can absorb some of the solar flux in the blue and UV parts of the spectrum.[21-23] This can be of importance especially when illumination is from the fullerene layer side, i.e. in non-inverted configurations. Other fullerene derivatives do have significant solar absorption (most notably those based on C70)[24] and can help generate additional photocurrent. It is therefore important to be able to compute optical properties of fullerenes, and that, in the aggregate state. Absorption properties of C60 and other fullerene derivatives in molecular form (solution) have been reported[17, 23] as well as approximate calculations in molecular solids.[19, 20] How much does the aggregate state affect the shape and the intensity of the absorption spectrum? This is not trivial to answer for any molecular solid; specifically for fullerenes, this question is further complicated by the fact that the absorption spectrum is not dominated by HOMO→LUMO transitions but involves contributions from transitions between many orbitals, necessitating the inclusion of *hundreds* states even in a single molecule calculation and proportionally more for molecular aggregates.[17] Such calculations would therefore be costly with TD-DFT (Time-Dependent Density Functional Theory).[25, 26] Further, TD-DFT is very sensitive to the exact shape of orbitals which contribute to the transitions and to errors in them.[27]

In this paper, we present a combined DFT – DFTB (Density Functional Tight Binding)[28] study of the effect of solid state packing on bandstructure and optical absorption of C60. We use DFT/TD-DFT as well as DFTB and its time-dependent extension, TD-DFTB,[29] to compute C60 in crystalline form, and to compute its optical properties. DFTB is an approximate DFT method that is about three orders of magnitude faster than DFT and is therefore attractive for molecular solid calculations due to its ability to treat large system sizes (such as C60 clusters considered here), and for optical property calculations due to the feasibility of considering large numbers of transitions in TD-DFTB, which is especially important for fullerenes. Is the quality of the orbitals computed with DFT(B) for aggregated fullerene molecules sufficient to produce a reasonable absorption spectrum?

We show that absorption spectra computed with TD-DFT (using different functionals) and TD-DFTB show redshift in C60 molecular clusters, which in the case of DFTB and DFT with a GGA (generalized gradient approximation) functional is unrealistically large. This is on top of TD-DFT's sensitivity to the approximation of the exchange and correlation functional which results e.g. in underestimated excitation energies with GGA. This follows directly from linear response TD-DFT's strong dependence on orbital energies and shapes, suggesting that the quality of the orbitals, which are delocalized by DFT (and DFTB) over more than one molecular unit, is insufficient to produce a quality spectrum with TD-DFT.

In solid state calculations, the dipole approximation has been popular[20, 30] as it is well amenable to periodic calculations. In it, the absorption spectrum is computed based on the imaginary part of the frequency dependent dielectric function $\varepsilon(\omega)$. Although a different ansatz, the dipole approximation, similarly to TD-DFT, also relies on orbital energies and shapes and critically depends on integrals over overlapping orbitals. We show that in the case of C60 aggregates it also leads to unrealistically large redshift.

In an attempt to overcome the shortcomings of both TD-DFT and the dipole approximation, we test an alternative approach in which we estimate the real polarizability from which the real and then imaginary part of $\varepsilon(\omega)$ and the spectrum are computed. We show that this approach (i) is less dependent on the choice of exchange-correlation functional, specifically, the redshift of the spectrum with PBE compared to B3LYP is only about 10% of the excitation energy (compared to about 20% with TD-DFT) and (ii) appears to produce much more reasonable spectra of molecular aggregates than GGA TD-DFT or the dipole approximation, compared to available experimental data. The approach is also found to be costlier than TD-DFT but on the other hand, it is perfectly parallelizable.

## 2 Computational Methods

The DFT calculations on molecules and clusters were performed using Gaussian 09.[31] The PBE,[32] B3LYP,[33, 34] and $\omega$B97XD[35] functionals with the LanL2DZ[36] basis set were used. The HOMO and LUMO energies obtained with LanL2DZ were compared to those obtained with larger basis sets 6-311g and 6-311++g(2d,2p) to ensure that the basis is appropriate. The absorption spectra were computed using TD-DFT[26] considering 150, 150 and 100 lowest singlet-singlet transitions for C60 monomer, dimer, and tetramer, respectively. TD-DFT, which implements Casida equations,[25] is extremely sensitive to the shape and localization of orbitals. The excitation spectrum $\omega$ is obtained from the eigenvalue problem

$$\begin{bmatrix} A & B \\ B & A \end{bmatrix} \begin{bmatrix} X \\ Y \end{bmatrix} = \omega \begin{bmatrix} -1 & 0 \\ 0 & -1 \end{bmatrix} \begin{bmatrix} X \\ Y \end{bmatrix} \qquad (1)$$

The elements of matrices *A* and *B* depend on the integrals

$$K_{ia\mu,jb\nu} = \iint \phi_{i\mu}^*(\boldsymbol{r})\phi_{a\mu}(\boldsymbol{r})\left(\frac{1}{|\boldsymbol{r}-\boldsymbol{r}'|} + \frac{\delta^2 E_{XC}}{\delta\rho_\mu(\boldsymbol{r})\delta\rho_\nu(\boldsymbol{r}')}\right)\phi_{j\nu}(\boldsymbol{r}')\phi_{b\nu}^*(\boldsymbol{r}')d\boldsymbol{r}d\boldsymbol{r}'$$

(2)

where indices *i, j* and *a, b* label occupied and virtual orbitals $\phi$, respectively, and indices $\mu$ and $\nu$ denote spin, $\rho$ is the density, and $E_{XC}$ the exchange-correlation energy.[25] Eq. 2 is very sensitive to the quality of the orbitals and to any errors in the orbitals, in particular, because it involves overlap integrals with a kernel. For example, the effect of errors in orbitals is much stronger on Eq. 2 than it is on orbital energies.[27]

Periodic DFT calculations of the C60 molecules and crystal were performed using SIESTA[37] with the PBE[32] functional. A DZP (double-ζ polarized) basis set was used, and the density was expanded in plane waves with a cutoff frequency of 150 Ry. Grimme D2 type dispersion corrections were used to simulate the molecular crystal.[38] Structures were optimized until forces on all atoms were below 0.02 eV/Å and stress (for solid state calculations) below 0.01 GPa. In crystal structure calculations, the Brillouin zone was sampled with 2×2×2 Monkhorst-Pack[39] *k* points. Molecular and cluster calculations were performed at the Γ point in a 20×20×20 Å cell for molecules and 20×20×30 Å for dimers (whose axis was along the z axis).

DFTB calculations were performed employing the self-consistent charge density functional scheme (SCC-DFTB)[28] using the DFTB+ 1.3 code.[40] SCC-DFTB is an approximate DFT approach derived from a second-order expansion of energy obtained by DFT. This method has been shown to provide near-DFT accuracy for systems for which it is has been parameterized. One of the best parametrized sets for DFTB for organic materials is the set (of Slater-Koster files) 3ob-3-1 with DFTB-3 capability.[41] This parameter set was benchmarked for organic systems and shows good accuracy.[42, 43] The 3ob-3-1 parameter set is used in this study with dispersion corrections using the Grimme parametres.[44, 45] For crystals, the Brillouin zone was sampled with 3×3×3 *k*-points. For the light absorption properties computed with TD-DFTB,[29] 300, 600, 800, and 200 excited states were considered for C60 monomer, dimer, tetramer, and octamer, respectively.

The states' excitation energies $E_i^{exc}$ and oscillator strengths $f_i$ obtained with TD-DFT and TD-DFTB were used to calculate the molar absorptivity $\mu$ as a continuous function of the excitation energy $E$ using

$$\mu = \frac{1.35 \times 10^4}{\sigma} \sum_i f_i \, exp\left[-2.772\left(\frac{E - E_i^{exc}}{2\sigma}\right)^2\right]$$

(3)

with the HWHM (half width half maximum) broadening $\sigma = 0.25$ eV.

The molecular polarizability $\alpha(\omega)$ was computed as a function of excitation frequency $\omega$ with DFT[46] in Gaussian 09, for monomers (with PBE and B3LYP functionals) and dimers (with PBE). Due to a high CPU cost and convergence issues at high values of $\omega$ with DFT, calculations of $\alpha(\omega)$ for tetramers (as well as dimers) were done using the semi-empirical PM6 method,[47] which allowed us to compare absorption spectra between a dimer and a tetramer. That is, we compare monomer to dimer in DFT, and dimer to tetramer in PM6 to estimate redshifts with increasing cluster size. The real part of the dielectric constant $\epsilon_r$ was computed from the molecular polarizability $\alpha(\omega)$ using the Clausius–Mossotti relation[48, 49]

$$\frac{\epsilon_r - 1}{\epsilon_r + 2} = \frac{N\alpha}{3\epsilon_0}$$

(4)

where $N$ is the numbers density of molecules and $\epsilon_0$ the permittivity of vacuum. The Clausius–Mossotti relation makes the so-called Lorentz local field approximation i.e. it is based on the assumption that the long-range interactions are isotropic and that there is no charge transfer between molecules, which is a reasonable approximation for neutral C60 in the ground state, as is confirmed by numeric results below. Eq. 4 has been used to compute real dielectric constant for optical material.[50]

The imaginary part of the frequency-dependent dielectric function $\epsilon_i(\omega)$ was computed with DFT in SIESTA using

$$\epsilon_i(\omega) = \frac{2e^2\pi}{\Omega\epsilon_0} \sum_{k,v,c} |\langle\psi_k^c|\hat{q}\cdot r|\psi_k^v\rangle|^2 \delta(E_k^c - E_k^v - \hbar\omega)$$

(5)

where $\Omega$ is the cell volume, indices $v$ and $c$ scan occupied and unoccupied $\psi_k^{c,v}$ orbitals (whose eigenstates are $E_k^{c,v}$), respectively, $k$ is the wavevector, and $q$ is the photon polarization vector. This is the dipole approximation, and the calculation is done in the "polycrystal" regime effectively averaging over $q$. The calculation of $\epsilon_i(\omega)$, which has been used in the literature to compute spectra of fullerenes,[20] therefore also critically relies on the shape of Kohn-Sham orbitals due to an overlap integral with a kernel and is very sensitive to errors / approximations. The SIESTA calculations necessarily use a non-hybrid functional (PBE). Comparison to Gaussian calculations of $\epsilon(\omega)$, for which hybrid functionals can be used, also allows us to study the effect of the choice of the functional.

The real and imaginary parts of $\epsilon(\omega)$ computed with these two methods (i.e. either Eq. 4 or Eq. 5) were used to compute, respectively, the imaginary and real parts using the Kramers-Kronig relations[51, 52]

$$\epsilon_i = \frac{2}{\pi} P \int_0^\infty \frac{\omega' \epsilon_r(\omega')}{\omega^2 - \omega'^2} d\omega'$$

(6)

$$\epsilon_r = -\frac{2}{\pi} P \int_0^\infty \frac{\omega' \epsilon_i(\omega')}{\omega^2 - \omega'^2} d\omega'$$

where P stands for the principal value.[53] The absorption spectrum (molar absorptivity $\mu$) is then computed as

$$M\mu(\omega) = \frac{\sqrt{2}\omega}{c}\left(\sqrt{\epsilon_r(\omega)^2 + \epsilon_i(\omega)^2} - \epsilon_r(\omega)\right)^{\frac{1}{2}}$$

(6)

where *M* is molar concentration and *c* the speed of light. The molar concentration is assumed to be that of the C60 crystal (i.e. even when computing the spectrum from single-molecule calculations), to highlight effects of molecular aggregation.

# 3  Results and discussion

## 3.1  Structures, bandstructures, and effect of aggregate state on bandstructure

FIG. 1 shows the crystal structure of C60 optimized with DFT in SIESTA. The initial structure (*fcc*-like with four molecular units per unit cell) was taken from Ref.[54] The structure optimized with DFTB is visually similar. The lattice constant obtained with both DFT in SIESTA and DFTB+ is 13.8 Å. FIG. 1 also shows clusters of two, four, and eight units cut out of the crystal structure. The densities of states (DOS) around HOMO/VBM and LUMO/CBM states are shown in FIG. 2, where the DOS of a single molecule, clusters, and the solid computed with different methods are compared. In ref.[17], we already established that the solid state leads to differences in energies between VBM/CBM of the solid and HOMO/LUMO of individual molecules on the order of 0.1 eV; here, we can see that the effect of the aggregate state on the DOS of C60 is well reproduced with about eight molecular units. This can have a significant effect on charge separation in donor-acceptor pairs with small driving force to charge separation, as a change of 0.1 eV in the driving force can change the separation rate by about a factor of two.[55] On the other hand, this magnitude of change in HOMO/VBM, LUMO/CBM, and the bandgap is not expected to have a major effect on light absorption spectrum at a single-molecule level (through changes in excitation energies of transitions); however, optical properties could be affected by changes in the molecular environment, and this is studied next.

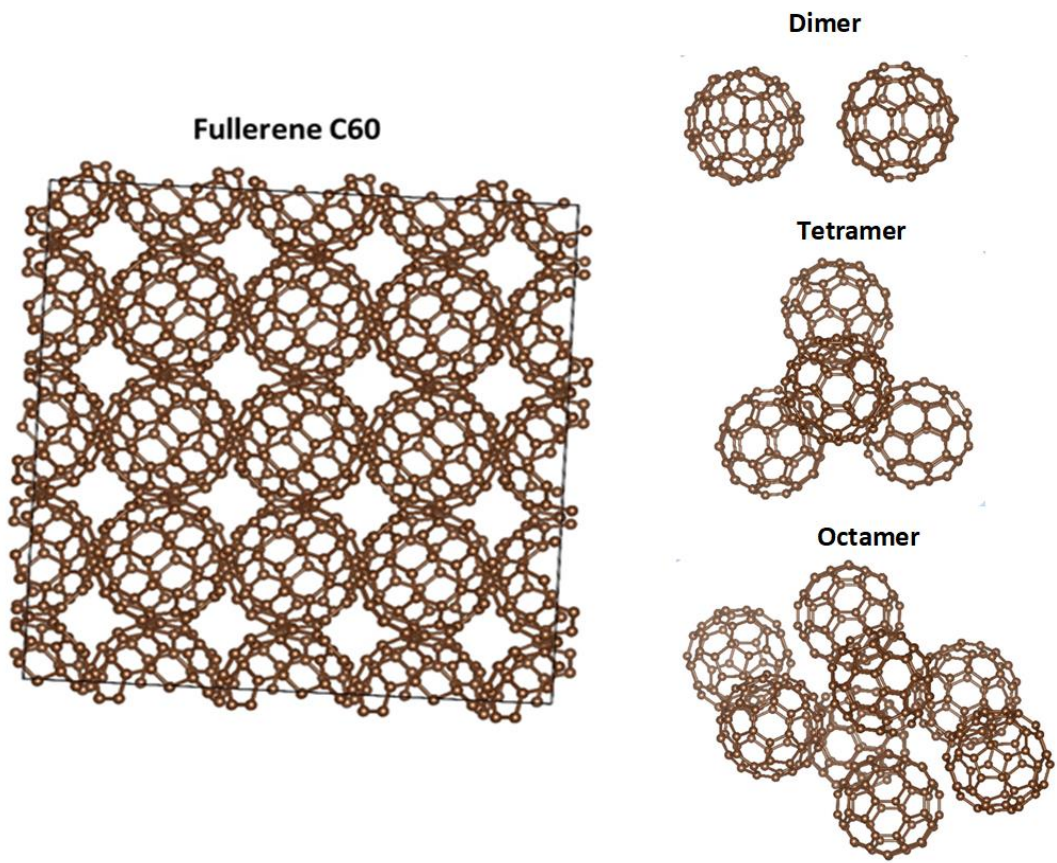

**FIG. 1.** The crystal structure of C60 and clusters with increasing numbers of units cut out from the crystal structure.

### 3.2 Visible absorption computed with linear response TD-TDFT

*TD-DFTB vs TD-DFT for single molecule absorption spectrum*

We first compute and compare TD-DFT and TD-DFTB absorption spectra, in that the former is expected to be quantitatively accurate[17] and can be used to benchmark the TD-DFTB spectrum. The TD-DFT and TD-DFTB absorption spectra of the C60 molecule are compared in FIG. 3. The excitation energies of absorption peaks appear to be underestimated by about 20% by TD-DFTB vs TD-DFT using the B3LYP functional. FIG. 3 also includes TD-DFT spectra computed with the PBE functional (on which the DFTB parameterization relies). We see that the absorption is also red-shifted vs PBE. This is largely due to redistribution of intensities among transitions rather than to lower energies of those transitions; indeed, the bandgap with DFTB (1.79 eV) is slightly larger than with PBE (1.74) and is smaller than with B3LYP (2.83 eV), as expected.

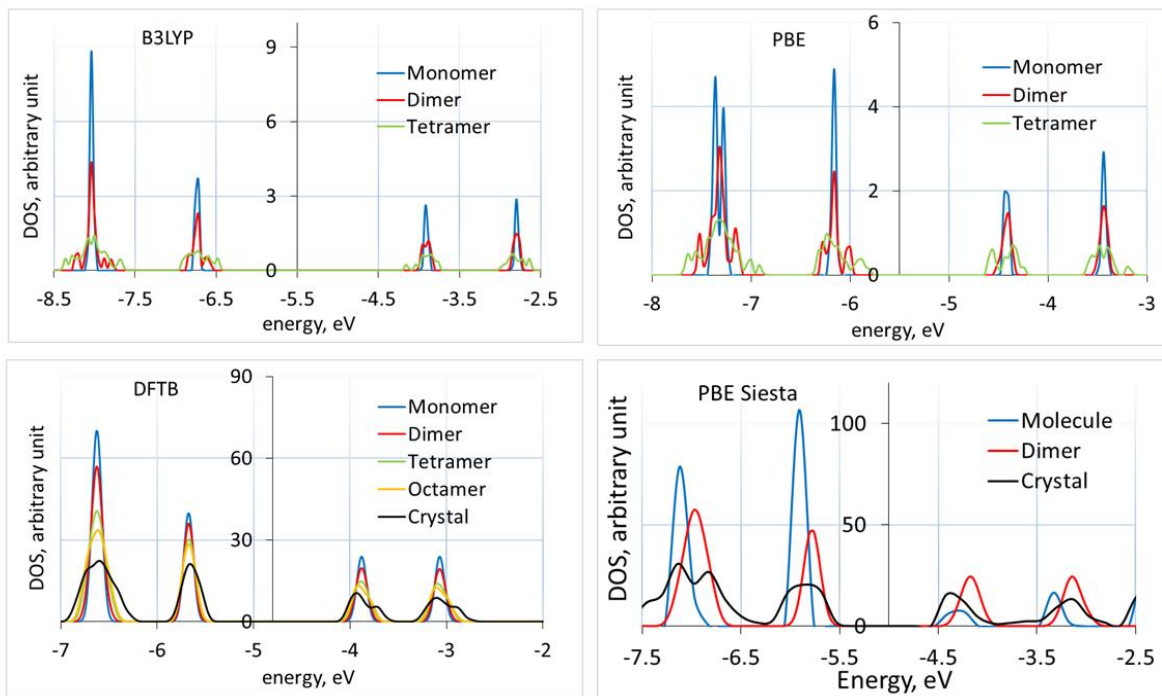

**FIG. 2.** The densities of states (DOS) of the C60 molecule, the clusters of different sizes, and crystal, computed with different methods. The ordinate axis is located between HOMO/VBM and LUMO/CBM.

*Effect of intermolecular interactions*

FIG. 3 also shows the TD-DFTB spectra computed on the clusters shown in FIG. 1. These spectra show a noticeable effect of molecular environment which mostly manifests itself in a red shift. The environment effect converges at about 8 molecular units; however, most of the effect is captured already at the dimer level. To confirm this conclusion at the TD-DFT level, we also computed absorption spectra of clusters (dimers and tetramers) with TD-DFT. In this case, to account for the fact that the hybrid-functional DFT is not expected to result in the same optimal interatomic arrangements as with dispersion-corrected DFTB, and to prevent effects due to stress, DFT calculations are performed by using DFT-optimized molecules at the same intermolecular distance as in the DFTB-optimized crystal. The results show that a similar effect of the presence of neighboring molecules is seen in both TD-DFTB and TD-DFT. We also computed the effect of changing monomer orientation at the same inter-molecular distance; the effect is relatively minor compared to the effect of the distance. Note that even though hundreds of transitions are included in TD-DFT and TD-DFTB calculations (see section 2), the spectrum of clusters does not extend much in energy due the very high density of transitions; this highlights difficulties of using TD-DFT to model effects of molecular aggregation.

The results of FIG. 3 imply a significant redshift of the absorption spectrum due to molecular aggregation. The redshift is so severe that it makes the spectra of clusters look unrealistic. Specifically, by comparing the experimental spectrum of C60 in Ref.[23] measured in a non-polar solvent, which can serve as an estimate of single-molecule absorption, to that of Ref.[56], which was measured in thin film, one observes that aggregation is expected to cause a mild redshift and an appearance of a shoulder (compare FIG. 1 of Ref.[23] with absorption onset around 3 eV and FIG. 2 of Ref.[56] which has an additional shoulder peaking at around 2.75 eV). The aggregation effect on the spectrum computed with B3LYP, where the spectrum of a single molecule is in semi-quantitative agreement with the experiment,[23] is relatively reasonable but appears exaggerated vs experiment, while that computed with PBE is unreasonable.[56] This is likely a method failure. TD-DFTB shows largely the same trend with cluster size as GGA TD-DFT, that is to say, this issue is not due to TD-DFTB but due to the underlying DFT approximation. This issue is addressed in the next section.

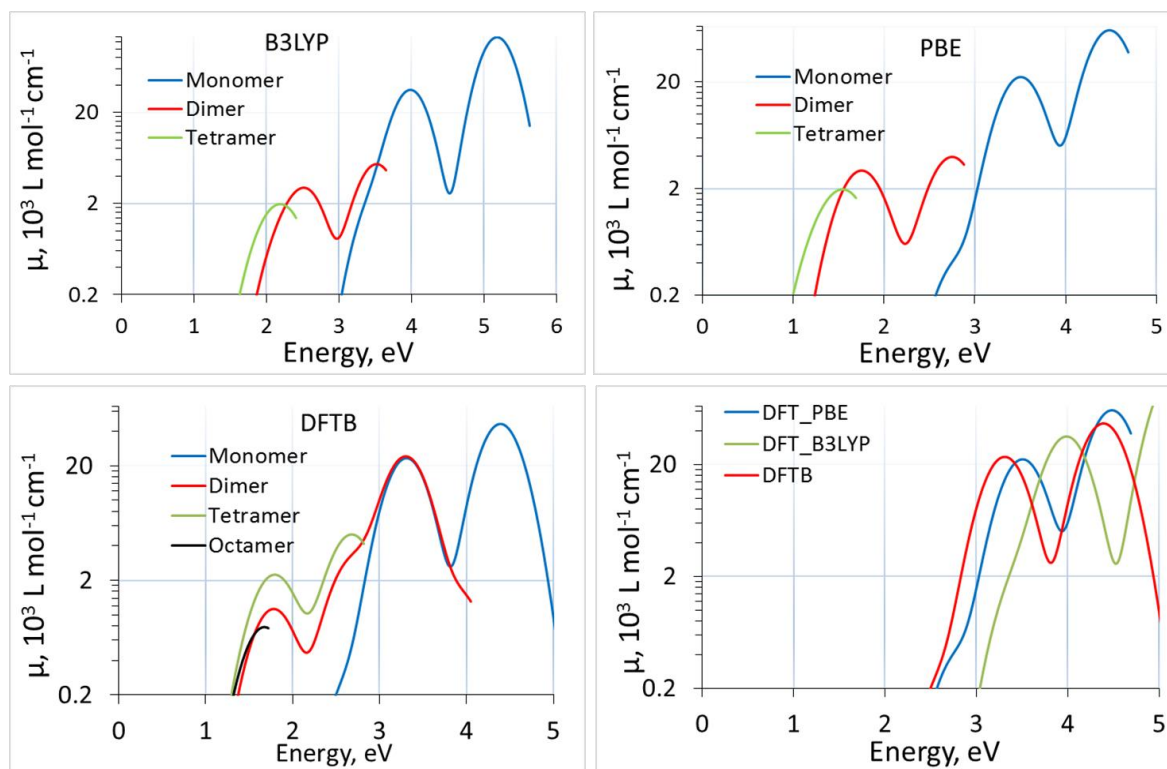

**FIG. 3.** Absorption spectra of C60 molecule and clusters computed with TD-DFT (PBE and B3LYP) and TD-DFTB. Note the logarithmic scale used here to better highlight the redshift due to aggregation.

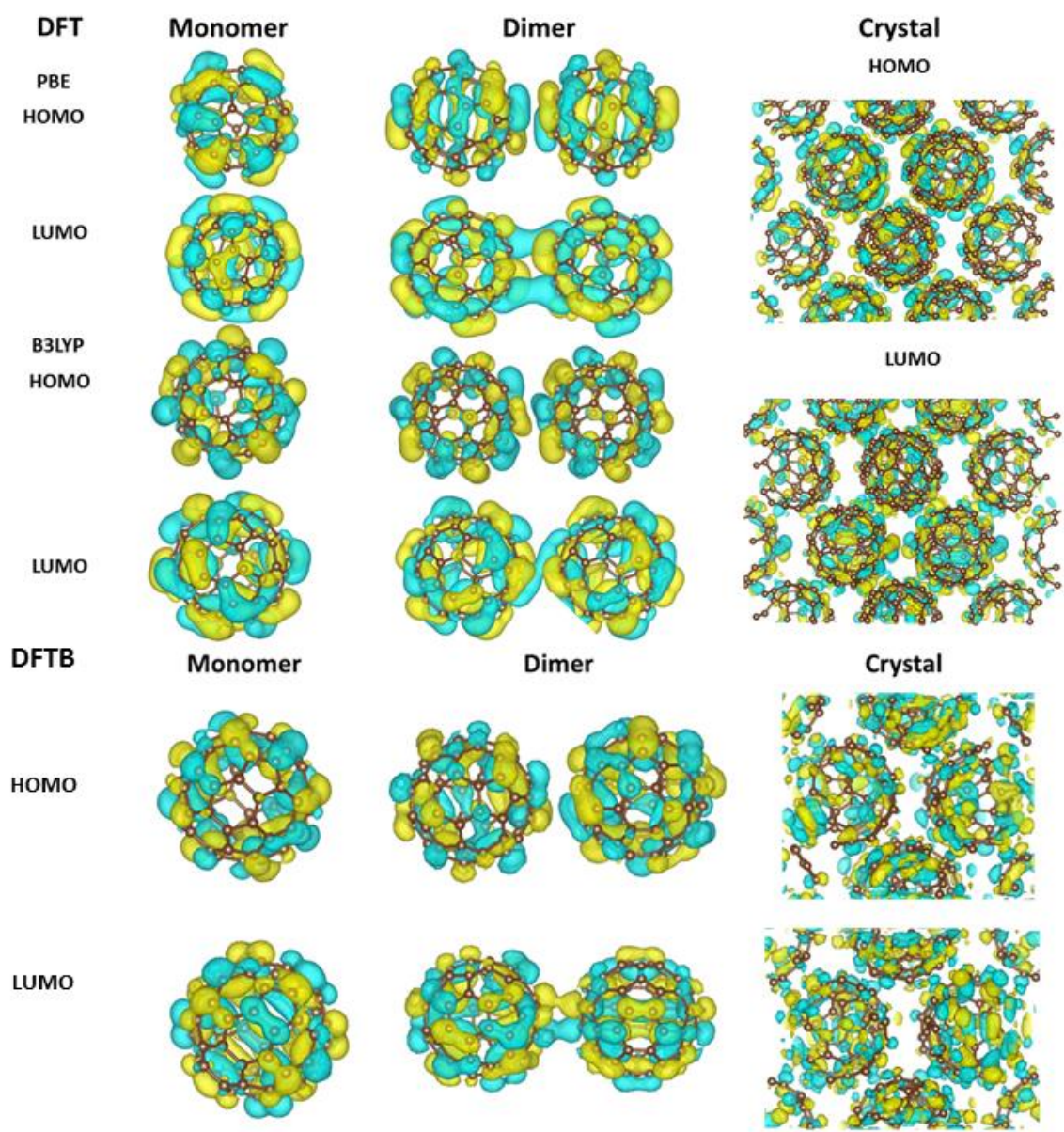

**FIG. 4.** Frontier orbitals of C60 molecule, clusters, and solid computed with DFT and DFTB with different functionals. The isovalue is 0.01 *a.u.* for molecules and dimers and 0.017 a.u. for the crystal.

*Orbital delocalization*

We have seen above that the effect of aggregation on the DOS is on the order of 0.1 eV (FIG. 2). The redshift observed in FIG. 3 is achieved not through a contraction of the gap, but through amplification of transition intensities of low-energy transitions. This has to do with orbital shapes. FIG. 4 shows frontier orbitals of a C60 molecule and clusters computed with DFTB and DFT with different functionals: GGA (PBE) and hybrid (B3LYP). They are delocalized over all molecular units; we have checked that delocalization persists even with the range-separated functional (ωB97XD). The orbitals are delocalized over neighboring molecular units with all these methods. Specifically, in periodic calculations, they are delocalized over all C60 units of the supercell. This is inconsistent with formation of small excitons in C60 and with hopping mechanism of electron transport which holds for C60.[57] Although it has been argued that nuclear vibrations lead to barrierless electron transfer,[58] they are not part of the present, 0K, model with which the TD-DFT spectra are computed. The delocalization therefore appears unphysical (i.e. not corresponding to the density distribution of real electrons). We note here that Kohn-Sham orbitals need not have physical meaning and need not have shapes equal to electron charge distribution of real photoexcited electrons; within the DFT formalism, Eqs. (1-2) are exact. However, as argued above, they are prone to strong sensitivity to the shape of orbitals and to any errors / approximations affecting them. An alternative route to compute the effect of molecular aggregation on the spectrum is therefore desirable, and this is considered next.

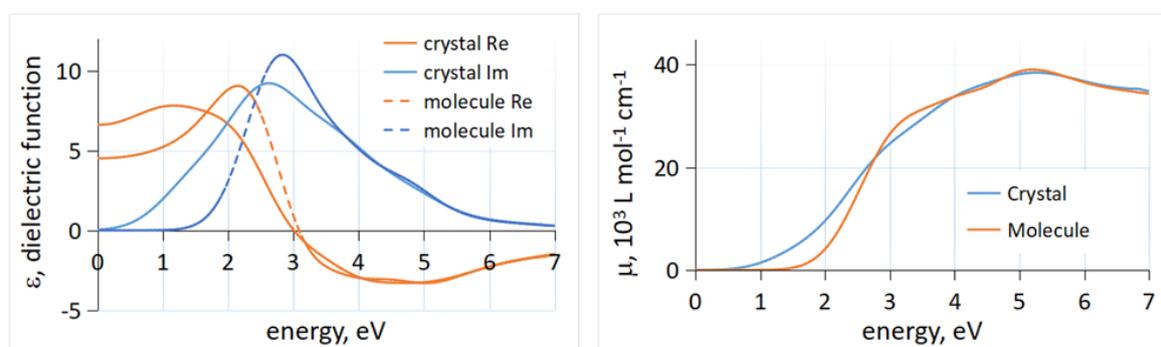

**FIG. 5.** Real and imaginary parts of the dielectric function (left) and absorption spectra (right) of C60 in molecular and crystalline state, where the imaginary part of the dielectric constant is computed by DFT in SIESTA. The dielectric function for the molecule is multiplied by the ratio of densities of molecules per simulation cell. The difference in the curves therefore shows the effect of aggregation.

### 3.3 Visible absorption computed from the frequency-dependent dielectric function

*Calculations of absorption spectra from the imaginary part of the dielectric function*

FIG. 5 shows the real and imaginary parts of the frequency dependent dielectric function and absorption spectra of the C60 crystal and C60 molecules computed based on the imaginary part of the dielectric function computed in SIESTA with the PBE functional. Similarly to TD-DFT, the spectrum computed with the GGA functional even for one molecule is severely redshifted (compared to the spectrum computed with B3LYP above) due to the dependence of the dipole approximation on the Kohn-Sham spectrum. In this calculation, one observes a similarly large redshift of the absorption spectrum due to aggregation as in the GGA TD-DFT calculation. The absorption coefficient that we obtain for the crystal is of similar magnitude as that computed in Ref.[20] with a similar approach (N.B. for comparison: that paper applied the scissor operator). The severe redshift, however, indicates that also with this method the effect due to aggregation is overstated. This is not surprising considering the critical dependence of Eq. 5 on the shape of the orbitals (overlap with a kernel), which makes this approach very sensitive to the approximations used. Therefore, the method of computing optical properties via Eq. 5, often used in solid state literature, is not accurate for C60 and likely for other fullerene based materials in solid state or other organic solids.

*Calculations of absorption spectra from the real polarizability*

FIG. 6 shows the real and imaginary parts of the frequency dependent dielectric function and absorption spectra of the C60 molecule computed based on the real part of the dielectric function computed from polarizability in Gaussian with the PBE and B3LYP functionals. The absorption spectrum shows the following remarkable features:

(i) Absorption peaks' energies computed with PBE and B3LYP functionals of the C60 molecules only differ by about 10%. While the spectrum computed with PBE is redshifted vs that computed with B3LYP, the amount of redshift is smaller than the typical underestimation of the excitation energies with PBE when using TD-DFT or the dipole approximation.[59] This is because this formalism,[46] although it does depend on integrals over occupied and unoccupied Kohn-Sham orbitals of the sort $\langle \phi_a \phi_j | \phi_b \phi_i \rangle$, $\langle \phi_a \phi_b | \phi_i \phi_j \rangle$,[46] does not depend on them in the same extremely sensitive way as TD-DFT or the dipole approximation in which the dependence (and therefore any errors) is amplified due to the kernel in the overlap integral. The spectrum computed with B3LYP is comparable with that computed with TD-DFT, which appears to be quantitatively accurate[17, 23] for both the onset of absorption (from about 3 eV) and magnitude of the extinction coefficient of the first peak around 4 eV (on the order of 20,000 L mol$^{-1}$ cm$^{-1}$).

(ii) Absorption spectra, computed with the same GGA functional, of the molecule and the dimer differ much less from each other than spectra computed with GGA TD-DFT and the dipole approximation, as is also shown in FIG. 6; there is also a relatively small change in the spectrum between the dimer and the tetramer. There is a small-intensity feature in the dimer spectrum (based on PBE) around 1.75 eV, but the main absorption peak is overall similar to that of the monomer, with a much more modest redshift and a shoulder appearing around 2.75 eV. This appears to be much more realistic than the large redshift obtained with TD-DFT and the dipole approximations (FIG. 3 and 5). Specifically, experimental spectra measured in non-polar solvents (expected to be comparable to single-molecule spectra computed in vacuum)[23] and in thin film[56] show similar change of spectral features due to aggregation, as explained in section 3.2.

This approach therefore appears to work well for C60. The cost of the calculation, however, was higher than that of TD-DFT, and we also faced convergence problems for tetramers at higher excitation energies. We also had to limit ourselves to a GGA functional for dimer calculations, and calculations of larger clusters were costly even with GGA. On the other hand, this approach is perfectly parallelizable, as $\epsilon_r(\omega)$ can be independently computed for each frequency.

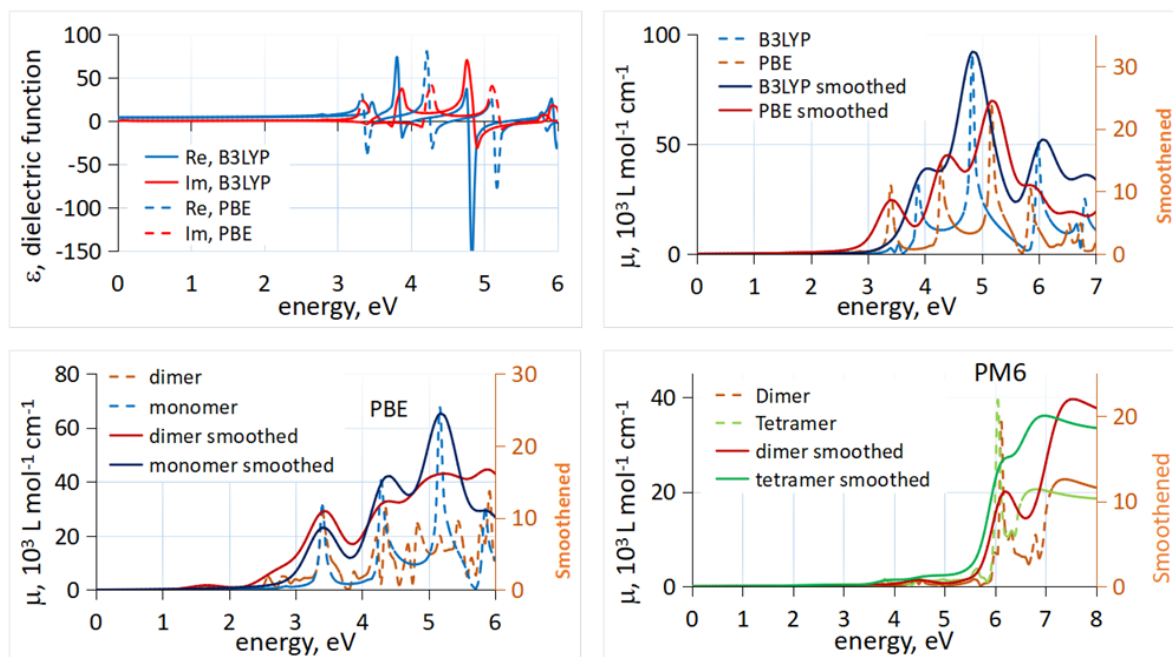

**FIG. 6.** Real and imaginary parts of the dielectric function and absorption spectra of C60 molecules and clusters, where the real part of the dielectric constant is estimated from polarizability computed by DFT in Gaussian with PBE and B3LYP functionals for monomers and dimers, as well as with PM6 for dimers and tetramers. On the plots of the spectra, the dashed curves follow from the Eqs. (4-5) and correspond to the left ordinate axes, and the smoothened curves are Gaussian-broadened with HWHM = 0.25 eV and correspond to the right ordinate axes.

## 4  Conclusions

We considered the effect of aggregation of C60 molecules on bandstructure and optical properties. We studied changes in bandstructure when going from single molecule to clusters of different sizes to solid. Aggregation causes contraction of the bandgap via stabilization of the LUMO and destabilization of the HOMO on the order of 0.1 eV. Clusters of eight C60 units mimic the effect of the solid on the bandstructure well. This effect was considered using DFT with a hybrid functional for clusters and a GGA functional for clusters and the solid as well as using DFTB for clusters and in solid state. All methods showed qualitatively similar results with expected errors in the gap due to the use of specific approximations. Dispersion-corrected DFTB predicted similar lattice parameters of solid C60 as dispersion corrected DFT. We previously showed that DFTB also provides accurate molecular structures of fullerenes.[17] DFTB can therefore be recommended as a fast and accurate method to model fullerene *structures* in aggregate state.

We then considered optical absorption spectra computed with (linear response) TD-DFT (using PBE and B3LYP functionals) and TD-DFTB. At the single molecule level, the spectra are qualitatively similar with quantitative differences which can be attributed to the use of specific functionals and parameterizations, i.e. single molecule spectra computed with all these methods can be practically useful as long as one accounts for unrealistic bandgap changes by using e.g. the scissor operator. However, spectra computed with TD-DFT(B) for C60 clusters showed significant redshifts vs single molecule; specifically, redshifts due to aggregation computed with a GGA functional and with DFTB were unrealistically large. The redshifts were stronger for larger clusters with convergence of the spectra around eight molecular units, but most of the effect is already seen at the dimer level. This is a qualitative error which can be attributed to the strong dependence of linear response TD-DFT on shapes and energies of Kohn-Sham orbitals, which makes it very sensitive to errors and approximations (such as those due the choice of the functional). The orbitals of clusters and in the periodic solid state were found to be delocalized over multiple (or infinite number of) C60 units, which may not reflect real spatial distributions of densities of photogenerated electrons and holes of this excitonic material.

We also computed the absorption spectrum from the imaginary part of the complex dielectric function using the popular dipole approximation. This approach was also found to lead to an unrealistic redshift upon aggregation, and much for the same reason of its critical dependence on Kohn-Sham orbital energies and shapes via overlap integrals with a kernel. Both TD-DFT and the dipole approximation therefore fail to account quantitatively for aggregation effects on the absorption spectrum. Their use is also complicated by the need to include multiple transitions as the absorption of fullerenes is not dominated by HOMO-to-LUMO transitions but transitions among many orbitals (very many in the case of clusters and solids).

We therefore considered an alternative approach, which appears to depend less critically on Kohn-Sham orbitals. We computed the real frequency-dependent polarizability, from which the real part of the complex dielectric function is computed, then the imaginary part, and finally the absorption spectrum. We find that this approach (i) does not suffer from severe underestimation of the excitation energies when using the cheaper GGA approximation and (ii) leads to a realistic change in the spectrum when C60 molecules aggregate, even with a GGA functional. The cost of the calculation was higher than that of TD-DFT, and we also faced convergence problems for tetramers at higher excitation energies; however, this approach is perfectly parallelizable, as $\epsilon_r(\omega)$ can be independently computed for each frequency. This therefore may be a promising approach to compute optical properties of organic crystals and clusters of molecules, and possibly of other types of materials. This also highlights the utility of developing methods to compute frequency dependent polarizability which would be more stable and less dependent on the quality of the Kohn-Sham spectrum and orbital shapes.

## Corresponding Author

*E-mail: mpemanzh@nus.edu.sg

## Conflicts of interest

There are no conflicts of interest to declare.

## Acknowledgments

This work was supported by the Ministry of Education of Singapore (AcRF Tier 1 grant). We thank Dr. Johann Lüder for assistance on parts on this work.